\pdfoutput=1
\documentclass[%
superscriptaddress,
nofootinbib,
twocolumn,
 amsmath,amssymb,
 aps,
prb,
]{revtex4-1}

\usepackage{graphicx}
\usepackage{chemarr}
\usepackage{times,longtable}
\usepackage{hyperref}
\usepackage{version}
\usepackage{color}
\usepackage{soul}
\usepackage{epsf,mathtools}
\usepackage[usenames,dvipsnames]{xcolor}

\graphicspath{{large/}}

\bibpunct{[}{]}{,}{n}{}{,}

\newcommand{\StochMatrixForm}{S_{\mu i}}
\newcommand{\StochMatrixFormJ}{S_{\mu j}}
\newcommand{\fluxI}{f_i}
\newcommand{\fluxImax}{f_i^{{\rm max}}}
\newcommand{\fluxImin}{f_i^{{\rm min}}}
\newcommand{\fluxJ}{f_j}
\newcommand{\EXT}{{\rm ext}}
\newcommand{\INT}{{\rm int}}
\newcommand{\muext}{{\rm \mu, \EXT}}
\newcommand{\muint}{{\rm \mu, \INT}}
\newcommand{\Vint}{V_\INT}
\newcommand{\Vext}{V_\EXT}
\newcommand{\GLC}{{\rm glc}}
\newcommand{\atp}{{\rm atp}}
\newcommand{\LAC}{{\rm lac}}
\newcommand{\CarbonDioxide}{\rm co2}
\newcommand{\OXYGEN}{\rm o2}
\newcommand{\OXPHOS}{{\rm ox}}

\newcommand{\LRA}{$\longleftrightarrow$}

\usepackage{chemarr}
\usepackage{epsf,mathtools}
\usepackage{graphicx}
\usepackage{dcolumn}
\usepackage{bm}

\begin{document}


\title[Inferring metabolic phenotypes from the exometabolome]{Inferring metabolic phenotypes from the exometabolome\\through a thermodynamic variational principle}

\author{Daniele De Martino}
\address{Center for Life Nano Science@Sapienza, Istituto Italiano di Tecnologia, Viale Regina Elena 291, 00161 Roma (Italy)}

\author{Fabrizio Capuani}
\affiliation{Dipartimento di Fisica, Sapienza Universit\`a di Roma, P.le A. Moro 2, 00185 Rome (Italy)}
\affiliation{CNR-IPCF, Unit\`a di Roma, Rome (Italy)}

\author{Andrea De Martino}
\affiliation{Dipartimento di Fisica, Sapienza Universit\`a di Roma, P.le A. Moro 2, 00185 Rome (Italy)}
\affiliation{CNR-IPCF, Unit\`a di Roma, Rome (Italy)}
\address{Center for Life Nano Science@Sapienza, Istituto Italiano di Tecnologia, Viale Regina Elena 291, 00161 Roma (Italy)}

\begin{abstract}
Networks of biochemical reactions, like cellular metabolic networks, are kept in non-equilibrium steady states by the  exchange fluxes connecting them to the environment. In most cases, feasible flux configurations can be derived from minimal mass-balance assumptions upon prescribing in- and out-take fluxes. Here we consider the problem of inferring intracellular flux patterns from extracellular {\it metabolite levels}. Resorting to a thermodynamic out of equilibrium variational principle to describe the network at steady state, we show that the switch from fermentative to oxidative phenotypes in cells can be characterized in terms of the glucose, lactate, oxygen and carbon dioxide concentrations. Results obtained for an exactly solvable toy model are fully recovered for a large scale reconstruction of human catabolism. Finally we argue that, in spite of the many approximations involved in the theory, available data for several human cell types are well described by the predicted phenotypic map of the problem.   
\end{abstract}

\maketitle

\section{Introduction}

The wealth of biological data acquired in recent years via high-throughput techniques has lead to increasingly refined descriptions of a cell's content in terms of  macromolecules and metabolites, as well as of the complex interactions between them that determine a cell's physiology.
Such interaction patterns represent the networks underlying cellular organisation. Of course, various such networks  (metabolic, signaling, protein-protein interaction, regulatory, etc.) operate on separate time- and space-scales, continuously cross-talking and reciprocally feeding into each other in single cells. In addition, because cells do not live in isolation, inter-cellular interaction networks can also be considered as representations of multicellular organization (e.g. tissues, organs, etc.). Physiologic functions emerge through the integration of collective biological interactions across different scales, from cellular-level, to tissue and organ level. While this  concept has always been at the center of biology, a formal and mathematical description of physiology in terms of the dynamics of network of networks has become possible only recently \cite{Bashan, Ivanov}. Obtaining a comprehensive picture of cellular activity from large-scale data is however still a major theoretical and computational challenge \cite{marx2013biology}. 

Mathematical network-based models represent a natural way to encode the inherent complexity of the interaction structure revealed by integrated biological data. Different approaches have been developed for a variety of cellular networks \cite{alon2006introduction}.  In many cases, the basic idea on which such schemes rely is that of optimization: assuming biological networks are optimized to perform a specific, context-dependent biological function (e.g. biomass or energy production in metabolic networks) one focuses on the network states that maximize the specific functionality. While such methods have met predictive and/or explanatory success \cite{Orth:2010if}, optimal states normally depend on a multitude of interaction parameters, which makes robustness an issue. Furthermore, identifying objective functions is not always straightforward, as testified e.g. by the variety of optimality criteria that can be used to describe, to different degrees, the same system \cite{Schuetz:2012p4035, Holzhutter:2004p4126}. In some cases, it is even hard to argue that something is being optimized at all. A different approach consists in placing the emphasis not on functional aspects but on the physical constraints under which such networks operate, and in trying to identify generic variational principles (similar to those derived for physical systems) by which the space of possible network states can be reduced to the `physically relevant' ones. Clearly, states selected by physical variational rules will in general be unable to pin down a specific biological functionality. The picture they provide will however be inherently robust and, in certain cases (specifically whenever physical constraints set the relevant limits to the network's operation), one may hope to obtain biologically relevant insight. 

One such case is possibly that of cellular metabolic networks. A cell's metabolism is, in essence, the complex network of enzyme-catalyzed reactions that processes nutrients to derive energy and molecular building blocks, while harvesting free energy from the environment and allocating it in the multiple tasks a cell has to accomplish. Metabolic network models have been extended to the scale of the whole genome and optimization-based schemes are routinely employed for their mathematical analysis  \cite{terzer2009genome,palssonbook}. Stationarity can be argued to be an appropriate assumption in order to investigate the productive capabilities of the network in terms of output metabolites or maximal flow of specific reactions (including biomass production). Therefore, upon defining suitable objective functions, one may resort to frameworks such as Flux Balance Analysis (FBA) for their study \cite{Orth:2010if}. At the computational level, FBA is normally a linear programming (LP) problem, and it has been particularly successful in predicting the growth rate of microorganisms in batch cultures. In cases in which a clear objective function is lacking, as is typical in multicellular organisms, an unbiased sampling of the steady states can provide useful information on the  organization of reaction pathways  or on the design of experiments \cite{Price:2004p4298}. The practical feasibility of sampling algorithms on genome-scale systems is however still a concern.

It is therefore important to understand how physics, and thermodynamics in particular, constrains the solution space of metabolic networks. Recently, a simple and intuitive physical description of steady states based on a thermodynamic variational principle has been proposed, according to which reaction networks occurring in a given volume and exchanging compounds with a stationary environment (a `bath') tend to minimize the rate of decay of entropy production \cite{DeMartinoPLoSONE_2012}. The latter quantity can be written explicitly in terms of the stoichiometry of the reaction network, of intracellular reaction fluxes and of extracellular metabolite levels. Therefore, the rule allows in principle to explore the physically viable intracellular flux configurations upon changing the composition of the environment. 

In this article we shall apply this framework to infer viable steady-state flux patterns in the catabolism of human metabolic networks, with the goal of clarifying the extent to which physics accounts for the switch  in cellular energetic strategies from fermentative to oxidative phenotypes that is observed in many cell types \cite{VanderHeiden:2009p3807, DiazRuiz:2011p4260, wolfe2005acetate, Brooks:2009p4313}.  We will show that, perhaps surprisingly, thermodynamic principles alone predict the crossover between different metabolic phenotypes using a small number of `environmental' control parameters, such as the external glucose, lactate, oxygen and carbon dioxide levels. Indeed, despite the crude approximations on which the theory is based, experimental observations fall remarkably well within the derived scenario. These results ultimately suggest that a more thorough understanding of the physical and chemical constraints under which biological networks operate may provide us with much conceptual insight and possibly predictive power.

\section{Method}
\label{SEC:variationalprinciple}

Let $\mathbf{S}$ denote the stoichiometric matrix of a given metabolic network with $N$ reactions and $M$ metabolites, with $\StochMatrixForm$ the stoichiometric coefficient of  compound $\mu$ in reaction $i$.  
Upon neglecting the discrete nature of molecules, noise and spatial gradients, the mass balance equations for the concentration $c_\mu$ of each compound $\mu$ may be written in terms of the reaction fluxes $\{\fluxI\}$ as
\begin{equation}
\label{EQ:concentrations}
\dot{c}_\mu =\sum_{i=1}^N \StochMatrixForm \fluxI~~.
\end{equation}
The Gibbs energy of reaction $i$ may in turn be decomposed in terms of the chemical potentials $g_\mu$ of the different compounds, i.e. 
\begin{equation}
\label{GibbsFE}
\Delta G_i = \sum_{\mu=1}^M \StochMatrixForm g_\mu 
\end{equation}
where, for a well-mixed and diluted system, the chemical potentials at constant pressure and temperature are given by
\begin{equation}
\label{chem_pot_expansion}
g_\mu = g_\mu^0+RT\log(c_\mu)~~,
\end{equation}
where $R$ is the ideal gas constant and concentrations are assumed to be measured in units of a fixed reference level.

Differentiating \eqref{GibbsFE} with respect to time and applying \eqref{chem_pot_expansion} and \eqref{EQ:concentrations}, one sees that
\begin{multline}
\frac{d}{dt}\Delta G_i = \sum_{\mu=1} \StochMatrixForm \dot{g}_\mu =RT \sum_{\mu=1}^M \StochMatrixForm \frac{\dot{c}_\mu}{c_\mu} = \\=RT \sum_{j=1}^N \fluxJ \sum_{\mu=1}^M \frac{\StochMatrixForm \StochMatrixFormJ}{c_\mu}~~.
\label{Hderivative_implicit}
\end{multline}
Introducing the shorthands
\begin{gather}
-\frac{\Delta G_i}{RT}= y_i ~~,~~~~~ J_{ij} =  \sum_{\mu=1}^M \frac{\StochMatrixForm \StochMatrixFormJ}{c_\mu}~~,
\end{gather}
Equation \eqref{Hderivative_implicit} can be re-written as
\begin{equation}
\dot{y}_i = -\frac{\partial H}{\partial \fluxI}~~,
\end{equation}
where
\begin{equation}
H = \sum_{i,j=1}^N J_{ij}\fluxI \fluxJ ~~.
\end{equation}
This elementary derivation suggests that the time evolution of Gibbs energies follows the gradients of the quadratic function $H$ of the fluxes.
In turn, steady flux states correspond to the minima of $H$. (See \cite{DeMartinoPLoSONE_2012} for a more precise microscopic derivation.) Notice that $H\geq 0$ by construction. It can furthermore be seen \cite{DeMartinoPLoSONE_2012} that $H$ corresponds to the rate of entropy decay, i.e. $H=-\ddot{S}/R$, with $S$ the internal entropy of the system \cite{kondeprigo}.

Note that, in terms of the concentrations, $H$ takes the form
\begin{equation}
H = \sum_{\mu=1}^M \frac{\dot{c}_\mu^2}{c_\mu}~~.
\label{Hdefinition}
\end{equation}
It is convenient to distinguish the levels of intracellular compounds ($c_{\muint}$) from those of extracellular ones ($c_{\muext}$), so that
\begin{equation}
H = \sum_{\mu=1}^M \frac{\dot{c}_{\muext}^2}{c_{\muext}}+\sum_{\mu=1}^M \frac{\dot{c}_{\muint}^2}{c_{\muint}}~~.
\end{equation}
In turn, the variations of intracellular and extracellular concentrations are linked by the exchange fluxes $u_\mu$. If we single out the latter (assuming a positive sign for fluxes  entering the cell), we get
\begin{equation}
\label{Hcellular}
H= \epsilon^2 \sum_{\mu=1}^M  \frac{u_\mu^2}{c_{\muext}}+\sum_{\mu=1}^M \frac{(\sum_{i=1}^N \StochMatrixForm \fluxI + u_\mu)^2}{c_{\muint}}~~,
\end{equation} 
where $\epsilon$ is the ratio of intracellular and extracellular volumes: $\epsilon = \Vint/\Vext$.

A trivial solution to the $H$-minimization problem is obtained by taking vanishing fluxes $\fluxI$ and uptakes $u_\mu$, leading to $H=0$. We shall focus on solutions carrying non-vanishing fluxes, corresponding to non-equilibrium steady states. Since it is reasonable to think that the volume ratio $\epsilon$ will typically be small, the second term in \eqref{Hcellular} dominates $H$. In metabolic networks, however, the constraints provided with models are normally compatible with internal homeostasis, implying $\sum_i \StochMatrixForm \fluxI + u_\mu=0$. Therefore the $H$ is minimized by minimizing the first (exchange) term alone. In summary, for given extracellular levels $c_\muext$, the variational principle takes the form of the optimization problem
\begin{equation}
\label{EQ:optimproblem}
\min_{\{\fluxI\},\{u_\mu\}} \sum_{\mu=1}^M \frac{u^2_{\mu}}{c_\muext} ~~~~~  \text{subj. to } 
\begin{cases}
  \sum_i \StochMatrixForm \fluxI = - u_\mu \\
\fluxI \in [\fluxImin,\fluxImax] \\
u_\mu \in [u_\mu^\mathrm{min},u_\mu^\mathrm{max}]~~
\end{cases}~~.
\end{equation}
For sakes of simplicity, we shall henceforth re-define
\begin{equation}\label{acca}
H=\sum_{\mu=1}^M \frac{u^2_{\mu}}{c_\muext}~~.
\end{equation}
In short, the above problem amounts to the minimization of a positive definite quadratic convex function in a convex polytope defined by the conditions \eqref{EQ:optimproblem}.  Finding a solution requires in general polynomial time \cite{kozlov1980polynomial}. Sampling all solutions uniformly, on the other hand, can either be done exactly upon knowing the vertices of the polytope,  or by stochastic methods (e.g. Monte Carlo). The latter procedure is also polynomial when a Hit-and-Run Markov chain is employed \cite{Smith:1996p4127, Turcin:1971, Lovasz:1999p4121, Almaas:2004p4295, DeMartino:2013p4294}.

In the following, we shall explicitly solve the above problem for specific metabolic networks and compare its solutions both with empirical data and with solutions derived from another variational principle that is widely employed to describe out-of-equilibrium systems.

\section{Results}

\subsection{Minimal model for ATP production}
\label{SEC:minimal_model}
\begin{figure}
\centering
\includegraphics[width=0.5\textwidth]{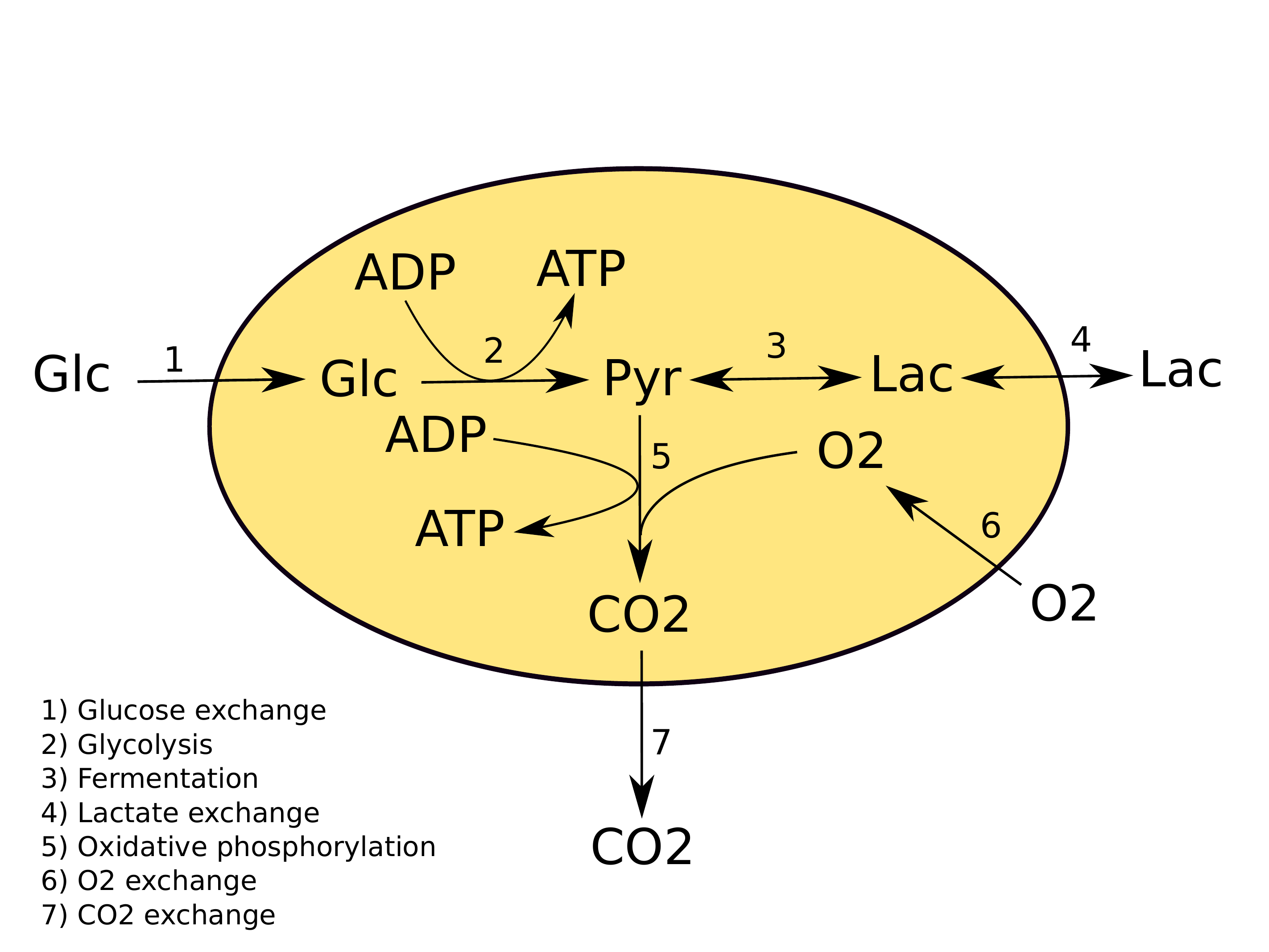}
\caption{Schematic representation of the reactions of the minimal model for ATP production. Reactions labeled 1, 4, 6, and 7 are exchange reactions of glucose (Glc), lactate (Lac), oxygen (O$_2$), and carbon dioxide (CO$_2$), respectively. Reaction 2 represents glycolysis with transformation of one molecule of glucose to two molecules of pyruvate and two ADPs to two ATPs. Reaction 3, which is reversible, transforms pyruvate to lactate and viceversa. Finally, reaction 5 represents oxidative phosphorylation where pyruvate and 3 molecules of oxygen are transformed to 3 molecules of carbon dioxide and 15 ATPs are created from 15ADPs. Note that stoichiometric coefficients are not explicitly reported in the network diagram.}
\label{FIG:simplemodel}
\end{figure}
As a first example, we consider a minimal model for the production of ATP from either glucose (through oxidative or fermentative metabolic pathways) or lactate (via oxidative pathways only), with the goal of inferring how a cell employs the different energy-producing pathways as a function of the extracellular levels of nutrients and waste products. 

Fig.~\ref{FIG:simplemodel} displays schematically the reactions of our minimal model. Each glucose molecule enters the cell and is transformed to 2 pyruvate molecules thereby converting two ADPs to ATPs (note that the stoichiometry is not depicted in the figure). Each pyruvate can either be used to produce further 15 ATPs, with the concomitant consumption of 3O$_2$ and production of 3CO$_2$ (oxidative pathway), or be transformed to lactate (fermentative pathway). The latter reaction is reversible so that lactate can be both expelled and intaken by the cell; in contrast, O$_2$ can only be intaken and CO$_2$ can only be expelled. We set the stoichiometry of this simple model to match the one of the complete network presented below; in general, the precise stoichiometry of ATP production via oxidation varies across species, with a yield in the range of 9 to 18 ATPs obtained per pyruvate molecule \cite{lenin}.

Denoting by $u_\mu$ the exchange reaction of metabolite $\mu$ (keeping in mind that $u^\mu>0$ for intakes and $u^\mu<0$ for outtakes), by $f_{\OXPHOS}$  the flux trough the oxidative pathway (labeled 5 in Fig.~\ref{FIG:simplemodel}), and by $f_{\atp}$ the ATP production flux (the sum of fluxes through reactions 2 and 5), the homeostatic intracellular steady state is defined by the equations
\begin{align}
\nonumber
f_{\atp} &= 2u_\GLC+15 f_{\OXPHOS} \\
\label{ss_simplemodel}
 f_{\OXPHOS} &= 2u_\GLC +u_{\LAC} ~~~~~~~~~,\\
\nonumber
-u_{\CarbonDioxide} &=u_{\OXYGEN}=3f_{\OXPHOS}
\end{align}
where the subscripts $\GLC$, $\LAC$, ${\CarbonDioxide}$, and ${\OXYGEN}$ stand for glucose, lactate, carbon dioxide, and molecular oxygen, respectively, and where we have used the fact that, by homeostasis, the flux through glycolysis equals $u_\GLC$. Since we have four extracellular species, $H$ as defined in \eqref{acca} is given by 
\begin{equation}
\label{H_simplemodel}
H=\frac{u_\GLC^2}{c_\GLC}+\frac{u_\LAC^2}{c_\LAC}+\frac{u_{\CarbonDioxide}^2}{c_{\CarbonDioxide}}+\frac{u_{\OXYGEN}^2}{c_{\OXYGEN}}~~.
\end{equation}
Fixing the ATP production flux conventionally to $f_{\atp}=1$ (this serves no specific purpose except fixing a scale for fluxes) and using the steady-state conditions \eqref{ss_simplemodel}, \eqref{H_simplemodel} can be re-cast as
\begin{equation}
H=\frac{(1-15f_{\OXPHOS})^2}{4c_\GLC}+\frac{(16f_{\OXPHOS}-1)^2}{c_\LAC}+9f_{\OXPHOS}^2(\frac{1}{c_{\CarbonDioxide}}+\frac{1}{c_{\OXYGEN}})~~.
\end{equation}
In turn, the value of $f_{\OXPHOS}$ that minimizes $H$ can be easily obtained by differentiating the above equation. The minimum of $H$ is obtained when
\begin{equation}
\label{EQ:foxminH}
f_{\OXPHOS} = \frac{\frac{15}{4c_\GLC}+\frac{16}{c_\LAC}}{\frac{15^2}{4c_\GLC}+\frac{16^2}{c_\LAC}+9(\frac{1}{c_{\CarbonDioxide}}+\frac{1}{c_{\OXYGEN}})}~~.
\end{equation}

Let us analyze how the emerging scenario changes with the extracellular levels. From the steady state conditions  \eqref{ss_simplemodel}, if there is no exchange of lactate, i.e. if $u_{\LAC}=0$, then glucose, which in this case is the only carbon source, is completely oxidized and $f_{\OXPHOS}=1/16$. Substituting this value in \eqref{EQ:foxminH}, we obtain that the glucose concentration corresponding to zero lactate exchange, which we denote as $c_\GLC^\star$, verifies
\begin{equation}
\label{g_threshold_SM}
c_\GLC^\star=\frac{5}{12} \frac{c_{\OXYGEN}c_{\CarbonDioxide}}{c_{\OXYGEN}+c_{\CarbonDioxide}}~~.
\end{equation}
Note that $c_\GLC^\star$ is independent of the external lactate concentration $c_\LAC$. For the physiological levels in the blood plasma ($c_{\CarbonDioxide}\simeq 30$ mmol, $c_{\OXYGEN}\simeq 5$ mmol), one finds $c_\GLC^\star \simeq 1.8$ mmol. On the other hand, combining \eqref{ss_simplemodel} with \eqref{EQ:foxminH}, and using expression (\ref{g_threshold_SM}), one finds that $u_\LAC \propto (c_\GLC^\star-c_\GLC)$. This suggests that \eqref{g_threshold_SM} defines a threshold separating, for any given levels of oxygen and carbon dioxide, different metabolic phenotypes. In particular, if  $c_\GLC>c_\GLC^\star$ one has $u_{\LAC}<0$ (there is a net lactate secretion), while $u_\LAC>0$ for $c_\GLC<c_\GLC^\star$ (corresponding to a net lactate intake).

We start by considering in detail the case $c_\GLC > c_\GLC^\star$. Here, ATP is produced using glucose exclusively, which can be channeled to both the oxidative and the fermentative pathway.
By recalling that one glucose produces two pyruvates, the fraction of oxidized glucose can be written as
$O = f_{\OXPHOS}/(2u_\GLC)\le1$, which, in the steady state described by \eqref{ss_simplemodel}, becomes simply
\begin{equation}
\label{EQ:OatSS}
O = \frac{f_{\OXPHOS}}{1-15f_{\OXPHOS}}~~.
\end{equation}
Substituting expression \eqref{EQ:foxminH} for $f_{\OXPHOS}$, one obtains the percentage of oxidized glucose as a function of the external metabolite levels, namely 
\begin{gather}
O = \frac{ay/x+1}{ay+1} ~~,
\end{gather}
where
\begin{gather}
a= \frac{15}{64} ~~~~~,~~~~~ y  = \frac{c_\LAC}{c_\GLC^\star} ~~~~~,~~~~~ x = \frac{c_\GLC}{c_\GLC^\star}~~.
\end{gather}
In the blood plasma, one has typically $c_\GLC\simeq 5$ mmol and $c_\LAC\simeq 1$ mmol, so that $O\simeq 0.92$. In other words, oxidation is the dominant energy-producing strategy for cells in contact with the blood.

If instead the glucose level is below threshold ($c_\GLC < c_\GLC^\star$), then there is a net lactate intake in addition to glucose and no fermentation is possible, as it would imply lactate secretion. In this case, the relevant quantity to consider is the fraction of  carbons that are intaken as glucose, or (recalling that one glucose molecule is converted to two pyruvate molecules) $G = {2u_\GLC}/({2u_\GLC+u_\LAC})\le 1$,
which at steady state is
\begin{equation}
\label{EQ:GatSS}
G = \frac{1-15f_{\OXPHOS}}{f_{\OXPHOS}}~~.
\end{equation}

\begin{figure*}
\centering
\includegraphics[width=\textwidth]{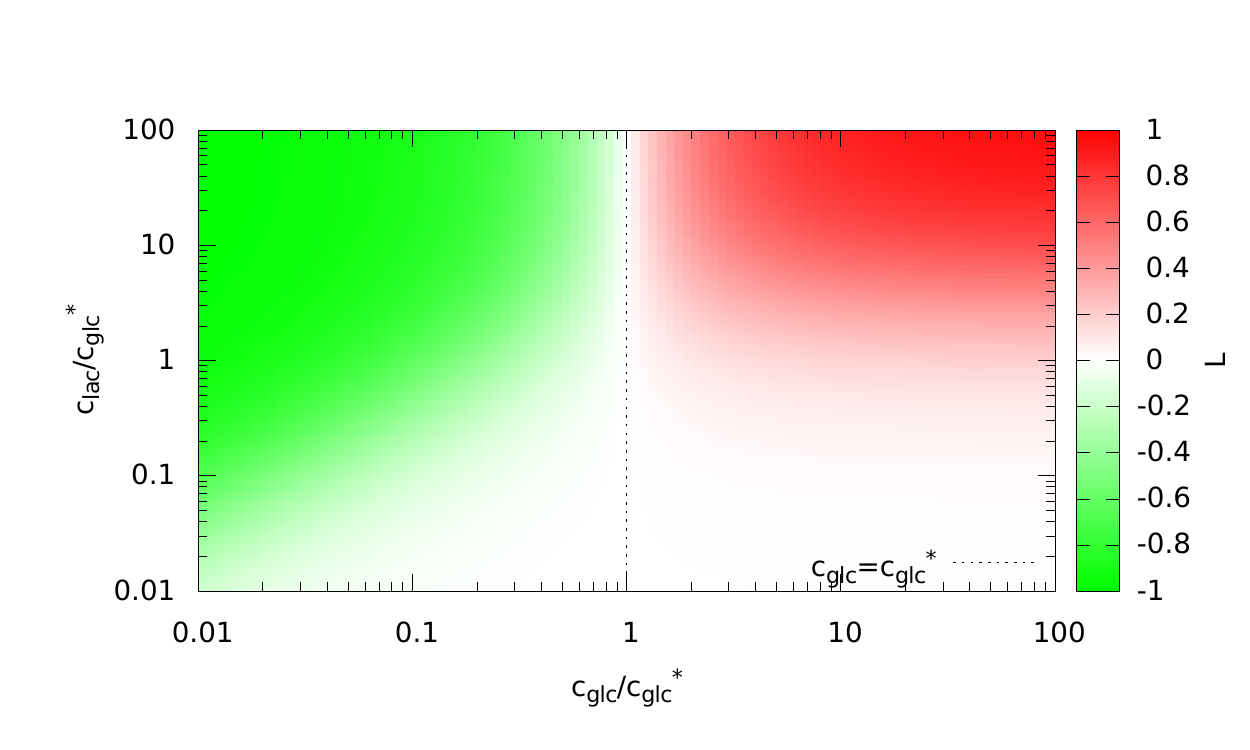}
\caption{Phenotypic map for the minimal model for ATP production. The critical value ${c_\GLC}={c_\GLC^*}$, defined in terms of the oxygen and carbon dioxide levels \eqref{g_threshold_SM}, separates the plane in two zones depending on whether cells intake (${c_\GLC}/{c_\GLC^*}<1$, $L<0$) or secrete (${c_\GLC}/{c_\GLC^*}>1$, $L>0$) lactate. 
The intensity of the colors represents the relative value of the lactate flux. For small values of ${c_\LAC}/{c_\GLC^*}$ the lactate flux is always negligible, while for ${c_\LAC}/{c_\GLC^*} \gtrsim10$ the amount of lactate exchange is large and, by crossing ${c_\GLC}/{c_\GLC^*}=1$, switches rapidly from a large lactate intake to a large lactate secretion.
L is defined in  \eqref{EQ:Lsimplemodel} and represents the fraction of carbon atoms exchanged as lactate.}
\label{FIG:simplecase}
\end{figure*}

To summarize: if $c_\GLC>c_\GLC^\star$ (or $c_\GLC/c_\GLC^\star>1$), the only source of carbon is glucose ($G=1$) and the fraction of carbons that are oxidatively processed is given by $O$. If instead $c_\GLC<c_\GLC^\star$ (or $c_\GLC/c_\GLC^\star<1$), oxidative carbon processing is the only possibility and $O=1$. The fraction of carbons intaken as glucose is concomitantly given by $G$. Noticing that $G = 1/O$, one can put different regimes together by defining a new function $L$, representing the normalized number of carbon atoms exchanged as lactate, being positive for outtakes and negative for intakes, which is given by 
\begin{equation}
L = \begin{cases}
G-1& \mbox{if } {c_\GLC}/{c_\GLC^\star}<1~~(O=1,~G<1)\\
1-O & \mbox{if } {c_\GLC}/{c_\GLC^\star}>1~~(O<1,~G=1)
 \end{cases}~~.
\label{EQ:Lsimplemodel}
\end{equation}
Note that  $L\in[-1,1]$. This allows us to organize possible metabolic phenotypes in a single diagram. The contour plot of $L$ in the plane $({c_\GLC}/{c_\GLC^\star},{c_\LAC}/{c_\GLC^\star})$ is displayed in Fig.~\ref{FIG:simplecase}, where four regions can be distinguished: a mainly fermentative regime with lactate outtake (red), a mainly oxidative regime with glucose as the exclusive carbon source (white, ${c_\GLC}/{c_\GLC^\star}>1$),
a purely oxidative regime with glucose as the main carbon source (white, ${c_\GLC}/{c_\GLC^\star}<1$),
and a purely oxidative regime with lactate as the main carbon source (green).

Within the $H$-minimization framework, the diagram maps out  the internal metabolic states of a cell as a function of the extracellular concentrations of glucose and lactate. The levels of oxygen and carbon dioxide are implicitly included via the parameter $c_\GLC^\star$ and serve as scaling factors. In essence, Fig.~\ref{FIG:simplecase} is the complete solution to the inference problem of determining the main carbon source and the pattern of pathway usage by a cell when the extracellular concentration of exchanged metabolites is known. We shall see that, despite its crudeness, the scheme just described captures the key features of cellular energetic strategies. Indeed, the same picture will emerge from a much more detailed model of cell metabolism.

\subsection{$H$-minimization versus Minimum entropy production}

It is instructive to compare these results with those obtained by minimizing entropy production, a well-known variational principle for biochemical systems, described e.g. in \cite{kondeprigo}. The general expression for the entropy production is given by \cite{kondeprigo}
\begin{equation}
\label{entropy_production}
T\dot{S}=\sum_{\mu=1}^M u_\mu g_\mu~~,
\end{equation}
where $T$ is the temperature and $g_{\mu}$ is the chemical potential of metabolite $\mu$. For the simple model discussed in the previous section, it takes the form
\begin{eqnarray}
\label{entropy_production_min_model}
T\dot{S} & = & \frac{1}{2} g_\GLC-g_{\LAC}+\left[ 3(g_{\OXYGEN}-g_{\CarbonDioxide})+16 g_{\LAC}-\frac{15}{2} g_\GLC\right]f_{\OXPHOS} \nonumber\\ 
& \equiv & \frac{1}{2} g_\GLC-g_{\LAC}+\alpha_{\OXPHOS}f_{\OXPHOS}~~.
\end{eqnarray}
In addition, the constraints $u_\GLC \geq 0$ (glucose is entering the cell) and $f_{\OXPHOS} \geq 0$ lead, via  \eqref{ss_simplemodel}, to $0 \leq f_{\OXPHOS} \leq 1/15$. Entropy production is easily seen to be minimized by two states only, depending on the sign of the coefficient $\alpha_{\OXPHOS}$. In particular, it is minimized by taking $f_{\OXPHOS}=0$ (complete fermentation) if $\alpha_{\OXPHOS}>0$, or $f_{\OXPHOS}=1/15$ (corresponding to no glucose uptake, and complete oxidation of lactate) if $\alpha_{\OXPHOS}<0$. 

Notice that, within this simple model, the minimization of entropy production predicts no intermediate (or mixed)  ATP producing strategy, and in particular no glucose oxidation. Therefore, according to this variational principle, fluxes will saturate their lower  or upper bounds depending on the sign of certain linear functions of the chemical potentials. Such a scenario inevitably leads to ``extreme'' regimes separated by sharp switches between them, which seems unlikely to agree with biological reality.

\subsection{Large-scale model of ATP production}

A realistic model of energy production by cells can be obtained by including the backbone of four ubiquitous pathways leading to ATP production, namely Glycolysis, Pentose Phosphate Pathway, Citric Acid Cycle, and Oxidative Phosphorylation. We have built such a network, extracting relevant reactions from the human reactome Recon-1 \cite{Duarte:2007p3938} (the complete list of reactions is reported in Table~\ref{TAB:reactions}). Altogether, the network comprises 49 chemical species and 45 reactions (18 of which are irreversible). The reactions include the shuttling of six metabolites: molecular oxygen, carbon dioxide, water, hydrogen, lactate, and glucose. Our goal is to analyze its feasible steady states according to the $H$-minimization principle, along the lines followed for the reduced toy model discussed above.

To apply the variational principle in this case,  we again impose steady state conditions for intracellular fluxes and single out the exchange reactions. Intracellular homeostasis is described by
\begin{equation}
\sum_{i=1}^N \StochMatrixForm \fluxI + u_\mu=0~~.
\end{equation}
These equations generate a large number of linear dependencies among fluxes, which can be resolved explicitly  by transforming the stoichiometric matrix to its Reduced Row Echelon Form (RREF) through Gaussian elimination \cite{Meyer:2000p4293}. It turns out to be possible to represent internal fluxes in terms of five degrees of freedom only, which we label as $(x_1,x_2,x_3,x,u)$. The RREF directly provides an expression of the different fluxes in terms of these parameters. Such expressions, emphasizing the biological meaning of the five independent degrees of freedom, are shown in the last column of Table~\ref{TAB:reactions}. In short, $x_1$ describes the so-called Rapoport-Luebering shunt (reactions catalyzed by DPGM and DPGase);  $x_2$ represents the ATP consumption, which will be fixed to 1 to match the corresponding production flux (see Sec.~\ref{SEC:minimal_model}); the flux through the Citric Acid Cycle is described by $x_3$; 
 $u$ corresponds to the glucose uptake; and, finally, $x$ represents the flux through the superoxyde dismutation that reduces O$_2^-$

\begin{table*}
\begin{tabular}{|l|l|l| }
\hline 
{\bf Enzyme}  & {\bf Reaction} & {\bf Steady state value}\\
\hline 
ACONT  &   CIT  \LRA ICIT  & $x_3$\\
\hline 
ACYP  &   13DPG + H$_2$O   $\longrightarrow$  3PG + H + Pi & $ -x_1-x_2 +x_3+233.1\bar{6}x +2u$\\
\hline 
AKGDm  &   AKG + CoA + NAD   $\longrightarrow$   CO$_2$ + NADH + SUCCoA & $x_3$ \\
\hline 
ATPS4m  &  ADP + 4 H + Pi   $\longrightarrow$  ATP + 3 H[M] + H$_2$O & $-x_3+250x$\\
\hline 
CSm  &  ACCoA + H$_2$O + OAA   $\longrightarrow$   CIT + CoA + H[M] & $x_3$\\
\hline 
CYOOm3  &  4 focytC + $7.92$ H[M] + O$_2$   $\longrightarrow$ 4 ficytC + 4 H + $1.96$ H$_2$O + $0.02$ O$_2^-$ & $50x$ \\
\hline 
CYOR{\textunderscore}u10m  &  2 ficytC + 2 H[M] + Q$_{10}$H$_2$   $\longrightarrow$ 2 focytC + 4 H + Q$_{10}$ & $ 100x $\\
\hline 
DPGM  &  13DPG   \LRA  23DPG + H  & $ x_1 $\\
\hline 
DPGase  &  23DPG + H$_2$O   $\longrightarrow$   3PG + Pi & $x_1$\\
\hline 
ENO  &  2PG  \LRA H$_2$O + PEP & $ x_3-16.8\bar{3}x+2u $\\
\hline 
FBA  &  FDP  \LRA  DHAP + G 3 P & $ x_3-16.8\bar{3}x+u $\\
\hline 
FUM  &  FUM + H$_2$O  \LRA MAL-L & $x_3$\\
\hline 
G6PDH2r  &  G6P + NADP  \LRA  6PGL + H + NADPH & $-3x_3+50.5x$\\
\hline 
GAPD  &  G3P + NAD + Pi   \LRA  13DPG + H + NADH & $x_3-16.8\bar{3}x+ 2u$\\
\hline 
GND  &  6PGC + NADP   $\longrightarrow$   CO$_2$ + NADPH + RU5P-D & $-3x_3+50.5x$\\
\hline 
HEX1  &  ATP + GLC   $\longrightarrow$   ADP + G6P + H & $ u $\\
\hline 
ICDHxm  &  ICIT + NAD  $\longrightarrow$   AKG + CO$_2$ + NADH & $ -5x_3+100x $\\
\hline 
ICDHy  &  ICIT + NADP   $\longrightarrow$   AKG + CO$_2$ + NADPH & $ 6 x_3 -100x $\\
\hline 
LDH  &  LAC-L + NAD   \LRA  H + NADH + PYR & $ 16.8\bar{3}x-2u $\\
\hline 
MDH  &  MAL-L + NAD   \LRA  H + NADH + OAA & $ x_3$\\
\hline 
NADH2{\textunderscore}u10m  &  5 H + NADH + Q$_{10}$   $\longrightarrow$   4 H + NAD + Q$_{10}$H$_2$ & $-x_3+100x $\\
\hline 
PDHm  &  CoA + NAD + PYR   $\longrightarrow$ ACCoA + CO$_2$ + NADH & $ x_3 $\\
\hline 
PFK  &  ATP + F6P   $\longrightarrow$ ADP + FDP + H & $x_3-16.8\bar{3}x+u $\\
\hline 
PGI  &  G6P  \LRA F6P & $ 3x_3-50.5x+u$\\
\hline 
PGK  &   13DPG + ADP  $\longrightarrow$ 3PG + ATP & $ x_2-250x $\\
\hline 
PGL  &  6PGL + H$_2$O   $\longrightarrow$  6PGC + H & $-3x_3+50.5x $\\
\hline 
PGM  &  2PG  \LRA  3PG & $-x_3+16.8\bar{3}x-2u $\\
\hline 
PYK  &  ADP + H + PEP   $\longrightarrow$ ATP + PYR & $x_3-16.8\bar{3}x+2u $\\
\hline 
RPE  &  RU5P-D   \LRA XU5P-D & $-2x_3+33.\bar{6}x $\\
\hline 
RPI  &   R5P  \LRA  RU5P-D & $ x_3-16.8\bar{3}x$\\
\hline 
SUCD1m  &  FAD + SUCC  \LRA FADH 2 + FUM & $ x_3$\\
\hline 
SUCOASm  &  ATP + CoA + SUCC   \LRA  ADP + Pi + SUCCoA & $ -x_3 $\\
\hline 
TALA  &  G3P + S7P   \LRA  E4P + F6P & $-x_3+16.8\bar{3}x $ \\
\hline 
TKT1  &  R5P + XU5P-D   \LRA  G3P + S7P & $-x_3+16.8\bar{3}x $\\
\hline 
TKT2  &  E4P + XU5P-D   \LRA F6P  + G3P & $-x_3+16.8\bar{3}x $\\
\hline 
TPI  &  DHAP   \LRA  G3P & $ x_3-16.8\bar{3}x+u$\\
\hline 
O2S reduction &  NADPH + O$_2^-$  + 2H   $\longrightarrow$  NADP + 2 H$_2$O & $x $\\
\hline 
FAD regeneration  &  Q$_{10}$ + FADH$_2$   $\longrightarrow$   Q$_{10}$H$_2$  + FAD  & $x_3 $\\
\hline 
ATP consumption &  ATP + H$_2$O   $\longrightarrow$   ADP + Pi + H & $ x_2$\\
\hline
Glucose exchange &  GLC \LRA  & $ u$\\
\hline
Lactate exchange &  LAC \LRA  & $ -2u+16.8\bar{3}x$\\
\hline
CO$_2$ exchange &  CO$_2$  \LRA & $ -50.5x$\\
\hline
O2 exchange &  O2 \LRA & $ 50x$\\
\hline 
\end{tabular}
\caption{Human catabolic reaction network. H[M] represents the (mitochondrial) hydrogen ion as an electromotive force, i.e. the protons transported across the inner mitochondrial matrix that give rise to the electrochemical gradient driving the ATPase. \label{TAB:reactions}}
\end{table*}

An expression for $H$ can be obtained directly through the expressions of exchange reactions, which are also defined by the steady state condition, as functions of the five independent degrees of freedom. The latter can be read off from Table~\ref{TAB:reactions}. In particular, the expressions for the exchange fluxes of lactate, glucose, carbon dioxide, and oxygen are given by
\begin{equation}
\begin{aligned}
u_{\CarbonDioxide} &= -50.5x \\ 
u_{\OXYGEN} &= 50 x \\ 
u_{\LAC} &= -2u+16.8\bar{3}x \\ 
u_{\GLC} &= u
\end{aligned}~~.
\label{EQ:udefinitions}
\end{equation}
Note that they depend on the two parameters $(x,u)$ exclusively. We neglect the exchanges of water and hydrogen ions assuming to work in  biochemical standard conditions, where the water level is taken to be large and hydrogen is buffered: in other words, $c_{{\rm h2o}}\gg 1$ and $\dot{c}_{{\rm h}}\simeq 0$. As a consequence, the corresponding terms in $H$ are negligible. Notice also that 
\begin{equation}
\frac{u_{\CarbonDioxide}}{6}+\frac{u_{\LAC}}{2}+u_{\GLC}=0~~,
\end{equation}
corresponding to the mass balance of carbon atoms. To characterize the domain of values of $(x_1,x_2,x_3,x,u)$ where the minimum of $H$ should be sought, one has to consider how the reversibility assignments encoded in the reaction network transfer to the independent variables. Indeed, while homeostasis defines linear dependencies among fluxes (i.e., linear equalities), the 18 irreversibility constraints (see Table~\ref{TAB:reactions}, with a positive (resp. negative) flux conventionally taken for the forward (resp. reverse) direction) define linear inequalities for the five remaining degrees of freedom. From Table~\ref{TAB:reactions} it can be easily recognized that all five degrees of freedom must be non-negative. A direct analysis of the redundancies of the constraints on the irreversible fluxes in Table~\ref{TAB:reactions} shows that the non-redundant constraints are the ones determined by the enzymes ACYP, ICIDHy, PFK, PGK, and PGL. In particular, one finds that the independent variables are cross-linked by the conditions
\begin{gather}
\begin{cases}
-x_1-1 +x_3+233.1\bar{6}x +2u \ge 0\\
x_3 -16.\bar{6}x \ge 0\\
x_3-16.8\bar{3}x+u \ge 0\\
1-250x \ge 0\\
-x_3+16.8\bar{3}x \ge 0, 
\end{cases}
\label{EQ:convexdomain}
\end{gather}
where we used the fact that $x_2=1$. Inequalities \eqref{EQ:convexdomain} together with the non-negativity constraints determine the convex domain where the minimum of H must be found.

Now, using \eqref{EQ:udefinitions}, $H$ is given by
\begin{equation}
\label{EQ:Hcompletemodel}
H = \frac{u^2}{c_{\GLC}}+\frac{(50x)^2}{c_{\OXYGEN}}+\frac{(50.5x)^2}{c_{\CarbonDioxide}}+\frac{(2u-16.8\bar{3}x)^2}{c_{\LAC}}~~.
\end{equation}
Note that $H$ is a quadratic function with absolute minimum in the origin. The latter point however lies outside the convex domain defined by  \eqref{EQ:convexdomain}. Therefore the feasible solution of the $H$-minimization problem will lie on an edge of the domain. Simple algebraic calculations reveal that $H$ is minimized on  the edge defined by
\begin{gather}
\begin{cases}
x_1=0 \\
x_3=16.8\bar{3}x \\
250x+2u = 1 \\
u\ge 0 \\
x \ge 0
\end{cases}~~.
\label{EQ:edge}
\end{gather}
The condition $250x+2u = 1$ reduces $H$ to a function of a single variable that, for convenience, we redefine to be $s=250x$ with $s\in[0,1]$. With this substitution, we are left with the simple problem of minimizing the function
\begin{gather}
H = \frac{(1-s)^2}{4c_{\GLC}}+\left[ \frac{1}{c_{\OXYGEN}}+\frac{1.01^2}{c_{\CarbonDioxide}} \right] \frac{s^2}{25}+\frac{(1-as)^2}{c_{\LAC}} ~~,
\end{gather} 
where $a=1+\frac{101}{1500}$. The minimum of H turns out to be attained when
\begin{equation}
s =  \frac{\frac{1}{4c_{\GLC}} + \frac{a}{c_{\LAC}}  }{ \frac{1}{4c_{\GLC}} + \frac{a^2}{c_{\LAC}} + \frac{1}{25} \left[ \frac{1}{c_{\OXYGEN}}+\frac{1.01^2}{c_{\CarbonDioxide}} \right]},
\label{EQ:minHfullmodel}
\end{equation}
which is a single point on the boundary of the convex domain.
\begin{figure*} 
\centering
\includegraphics[width=0.7\textwidth]{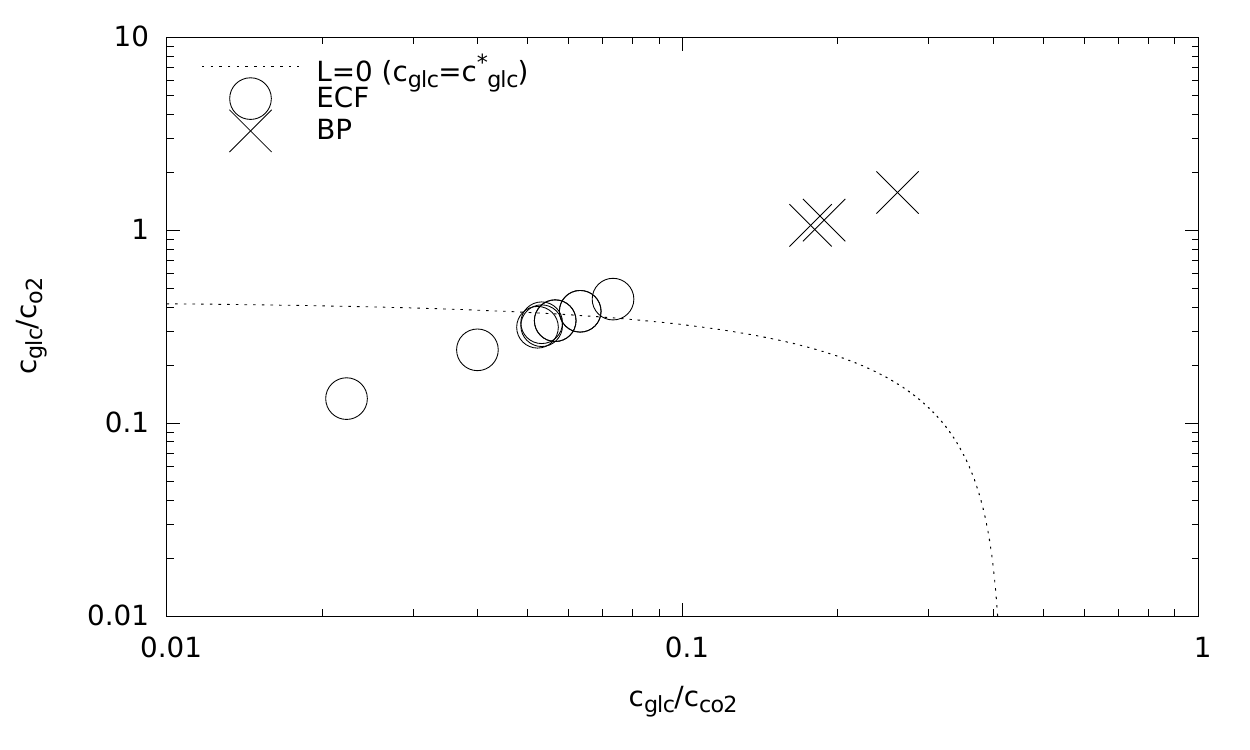}
\includegraphics[width=0.7\textwidth]{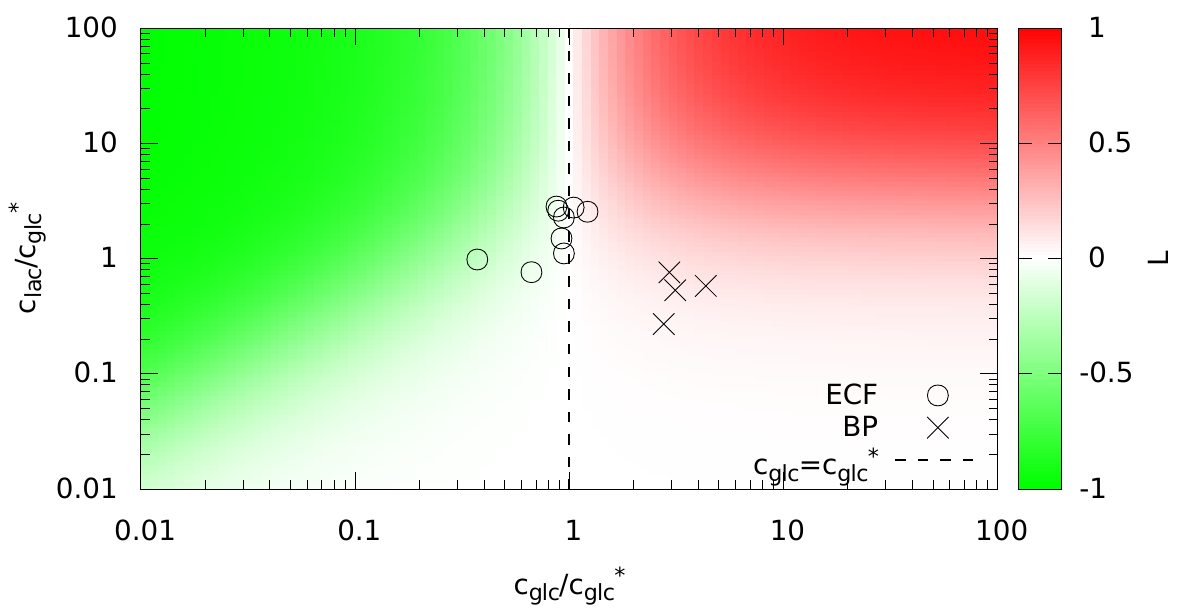}
\caption{Phenotypic map for the realistic model for ATP production.
Experimental measurements of glucose and lactate concentrations in extra cellular fluids (ECF, circles) and blood plasma (BP, crosses) are superimposed; we set $c_{\CarbonDioxide}=30$ mmol and $c_{\OXYGEN}=5$ mmol.
(a) critical line $c_{\GLC}=c_{\GLC}^*$ (dashed line) in the plane $({c_{\GLC}}/{c_{\CarbonDioxide}}, {c_{\GLC}}/{c_{\OXYGEN}})$. Below the line the phenotype is oxidative with partial intake of lactate, above the line the phenotype is partially fermentative with lactate outtake.  (b) Fraction of lactate uptake, with respect to glucose ($L<0$, green), or with respect to carbon dioxide ($L>0$, red) in the plane $({c_{\GLC}}/{c_{\GLC}^*}, {c_{\LAC}}/{c_{\GLC}^*})$. $L$  is defined in  \eqref{EQ:Lcompletemodel} and represents the fraction of carbon atoms exchanged as lactate. }
\label{FIG:completemodel}
\end{figure*}
As for the minimal, we are interested in describing the emerging metabolic phenotypes in terms of (a) the pattern of pathway utilization, and (b) the substrates from which ATP is preferentially produced.  The emerging scenario is indeed very similar to that derived in the simpler case. In specific,
\begin{itemize}
\item $s\to 0$ for $c_{\OXYGEN} \to 0$ or $c_{\CarbonDioxide} \to 0$, corresponding to glucose  intake and complete fermentation with lactate outtake;
\item $s\to 1/a$ for $c_{\LAC} \to 0$, corresponding to glucose intake and complete oxidation;
\item $s\to 1$ for $c_{\GLC} \to 0$, corresponding to lactate intake and complete oxidation.
\end{itemize}
In other words, as $s$ is changed in $[0,1/a]$ one passes continuously from fermentative ($s=0$) to oxidative ($s=1/a$) phenotypes, with lactate outtake, whereas for $s\in[1/a,1]$ one has complete oxidation without fermentation, and the preferred fuel switches continuously from glucose ($s=1/a$) to lactate ($s=1$). The value $s=1/a$ defines the curve 
\begin{equation}
\frac{240}{101} \left[ \frac{1}{c_{\OXYGEN}}+\frac{1.01^2}{c_{\CarbonDioxide}} \right]\equiv\frac{1}{c_{\GLC}^\star}~~,
\end{equation}
through which a threshold value $c_{\GLC}^\star$ for the glucose level $c_{\GLC}$ is obtained that, in parallel with the minimal model, can be conveniently used to separate different metabolic phenotypes.

In physiological conditions for the blood ($c_{\OXYGEN}\simeq 5$ mmol and $c_{\CarbonDioxide}\simeq 30$ mmol),  one has $c_{\GLC}^\star\simeq 1.8$ mmol. Below threshold, one finds partial lactate intake, whereas above threshold partial lactate outtake (fermentation) is observed. In this realistic model, the fraction of oxidized glucose is given by
\begin{equation}
\label{EQ:Ocompletemodel}
O=\frac{u_{\CarbonDioxide}}{u_{\CarbonDioxide} + 3u_\LAC}~~.
\end{equation}
For the average value of glucose and lactate in blood ($c_{\GLC}\simeq 5$ mmol and $c_{\LAC}\simeq 1$ mmol), \eqref{EQ:Ocompletemodel} predicts a small lactate outtake, the phenotype being almost completely oxidative (the fraction of glucose oxidized is $O\simeq 0.92$). In Fig.~\ref{FIG:completemodel} (top panel) we display the critical curve  $c_{\GLC}^\star$ in the plane $({c_{\GLC}}/{c_{\CarbonDioxide}}, {c_{\GLC}}/{c_{\OXYGEN}})$.

To make contact with empirical results, we have considered the experiments discussed in  \cite{zilberter2010neuronal, abi2002striking, harada1992cerebral}, where data are given for glucose and lactate levels for the Blood Plasma (BP) and for the Extra Cellular Fluid (ECF) in brain tissues, both in human and in rat. By plotting these data on the $({c_{\GLC}}/{c_{\CarbonDioxide}}, {c_{\GLC}}/{c_{\OXYGEN}})$ plane, we can in principle assess whether the metabolisms of brain cells in such environments is fermentative or oxidative and what is their preferred energy source. However, lacking data for carbon dioxide and oxygen levels, in order to carry out a detailed comparison we have employed the physiological values $c_{\OXYGEN}\simeq 5$ mmol and $c_{\CarbonDioxide}\simeq 30$ mmol. Strikingly, experimental points distribute so that $H$-minimization predicts a lactate intake for cells in the ECF and a lactate secretion for cells in the BP, in agreement with the conclusions drawn in \cite{zilberter2010neuronal}. 
\begin{table}
\centering
\begin{tabular}{|l|l|l|l|l|}
\hline 
Area  & $c_{\GLC}$ (mmol)  &  $c_{\LAC}$ (mmol)  & Lac in/out & lac $\%$  \\
\hline 
Human cortex BP  & 5.64 & 0.96 & OUT & 7.5 \\
\hline
Human cortex ECF  & 1.57 & 5.1 & IN & 6 \\
\hline
Rats hippocampus BP  & 7.84 & 1.05 & OUT & 9\\
\hline
Rats hippocampus ECF  & 1.66 & 2.7 & IN & 3 \\
\hline  
\end{tabular}
\caption{Inferred values of lactate uptake from experimental glucose and lactate levels in brain, for 
Humans \cite{abi2002striking} and Rats \cite{harada1992cerebral}. The first column reports the source of the experimental data. The second and third columns report the measured levels of glucose and lactate, respectively. Finally, the  last two columns show our estimate for the direction and of lactate and for the absolute value of $|L|$, respectively. $L$ is defined in  \eqref{EQ:Lcompletemodel} and represents the fraction of carbon atoms exchanged as lactate.}
\label{TAB:values}
\end{table}
Since Fig.~\ref{FIG:completemodel} (a) does not describe the dependence on lactate levels, it cannot provide the relative value of lactate uptake, with respect to glucose intake or carbon dioxide outtake. This information can however be retrieved by  studying how experimental points distribute in the $({c_{\GLC}}/{c_{\GLC}^\star}, {c_{\LAC}}/{c_{\GLC}^\star})$ plane with respect to the contour plot of the quantity $L$ (the normalized number of carbon atoms exchanged as lactate), which can be defined in analogy to \eqref{EQ:Lsimplemodel} as
\begin{equation}
L = \begin{cases}
\frac{u_{\LAC}}{2u_{\GLC} + u_\LAC} & \mbox{if } {c_\GLC}/{c_\GLC^\star}<1\\
-\frac{3u_{\LAC}}{u_{\CarbonDioxide} + 3u_\LAC} & \mbox{if } {c_\GLC}/{c_\GLC^\star}>1~~.
 \end{cases}
\label{EQ:Lcompletemodel}
\end{equation}
The resulting phenotypic map is shown in Fig.~\ref{FIG:completemodel} (b), where the emerging scenario is that of a partial fermentation in the BP and partial lactate intake in the ECF.  Values for the experimental data measured for both ECF and BP are reported in Table~\ref{TAB:values}.

\section{Discussion}

Metabolic flux analysis for cell-autonomous systems and for populations \cite{harcombe2014metabolic} is a rich, fruitful and expanding field of research, displaying multiple connections between theory and experiments. For the greatest part, modeling schemes rely on the possibility to identify objective functions by which the complexity of the space of possible solutions can be reduced, focusing on optimal flux patterns exclusively. In this way, given a set of uptake fluxes describing a cell's exchanges with the environment, the physiologically relevant states of the cell can be mapped onto a small set of flux configurations (possibly reduced to a single point), which can be studied in detail by different computational techniques. Sampling methods allow in principle to explore the solution space beyond optimality, to characterize, e.g., sub-optimal states, correlations, etc, and improving such techniques is one of the current frontiers of the field. At the other end of the modeling spectrum, one may be interested in characterizing how physics constrains the solution space by means of variational principles that characterize steady states in terms of the minima of specific functionals that are physically- (rather than biologically) motivated. The thermodynamics of reaction networks, a subject that goes back to at least \cite{oster1973network} and it has recently gained attention from a stochastic perspective
\cite{gaspard2004fluctuation,schmiedl2007stochastic}  suggests that steady states should be characterized in terms of their entropy production. Since such principles should hold for cellular systems as well \cite{beard2007relationship, polettini2014irreversible, DeLaFuente2013, kondeprigo}, it is interesting to study what type of information can be obtained about a cell's metabolism by applying these ideas. 

In this work, we have studied the problem of inferring the metabolic phenotype from the level of metabolites in the extracellular space (i.e. from the exometabolome) using a thermodynamic variational principle for the steady states of a chemical reaction network. The variational principle, which holds for slowly varying chemical potentials, amounts to the minimization of the rate of decay of entropy production. From a conceptual viewpoint, it merely allows us to select intracellular flux states that are compatible with the observed extracellular concentrations. From a technical viewpoint, it requires the minimization of a semi-positive definite quadratic function of the fluxes in the space of feasible steady states, and can be carried out in polynomial time. 

We have applied it to the catabolic core of a detailed genome-scale reconstruction of the human metabolic reactome with the goal of characterizing the conditions under which cells switch from a fermentative to an oxidative phenotype as a function of the external levels of key environmental indicators like glucose, lactate, oxygen and carbon dioxide. Our results indicate that cells transition from one phenotype to the other in a continuous, modulated way, and that mixed phenotypes are possible.
This is in line with empirical knowledge \cite{Molenaar:2009p3978, Barros:2013p4309}
and at odds with the scenario predicted by another widely used variational principle (that of minimal entropy production),
which predicts sharp transitions \cite{kondeprigo} rather than smooth cross-overs.
The scenario we obtain is also recovered in an exactly solvable toy model that only retains the main features of the key energy producing pathways.
Quite remarkably given the crudeness of the variational principle we employ, upon inferring from experimental values of glucose and lactate levels in the brain we find very moderate levels of lactate exchange, in specific a small intake in the extra-cellular fluids and a small outtake in the plasma. Experimental evidence compares well with the predicted ``phase structure'', suggesting that indeed fundamental physical considerations might suffice to explain at least part of the evidence on cellular energy production strategies. 
One of the limitations of the approach discussed here is that, in principle, knowledge of intracellular substrate levels (besides extracellular concentrations) is required to solve the full-fledged variational problem. In this work, we have circumvented this difficulty by assuming that a mass-balanced flux pattern for the intracellular state. This assumption is justified for the type of systems we consider, but cannot be expected to hold generically. It would be interesting to see how strongly solutions for large- or genome-scale networks depend on the particular internal metabolite pools. In turn, characterizing robustness to fluctuations in metabolite levels may highlight the presence of bottlenecks in the reaction network. On a more abstract level, it would also be important to generalize these considerations to a dynamical setting, e.g., characterizing trajectories (as opposed to steady states) in terms of physical or thermodynamical variational principles. 

\bibliographystyle{plos2009}
\bibliography{minH}

\end{document}